\documentstyle[epsfig,conf-X]{article}
\begin{document} 
\small
\heading{%
%
Galaxies Surveys with Future X-ray Observatories

\centerline{-- or--}

\noindent
How do we Build Future X-ray Observatories to Study Galaxies?
}
\par\medskip\noindent
\author{%
G. Fabbiano
}
\address{%
Harvard-Smithsonian Center for Astrophysics, 60 Garden St.,
Cambridge MA 02138, USA
}

\begin{abstract}

\noindent
Galaxies are key objects for the study of cosmology,
the life cycle of matter, and stellar evolution. X-ray observations
have given us a new key window into these building blocks of the 
Universe, that allows us to investigate their hot gaseous component.
While significant advances in our knowledge are expected from Chandra 
and XMM, there is a number of fundamental 
questions that require a next generation of X-ray telescopes.
These telescopes need to be 10-30 sqmeter class telescopes,
with Chandra-like resolution, and with a suite of instruments
allowing spectral imaging at moderate to high resolution.

\end{abstract}
\section{Why Galaxies in X-rays?}

Galaxies are key objects for the study of cosmology, the life
cycle of matter, and stellar evolution. Insights into the nature
and evolution of the Universe have been gained by using galaxies 
to trace the distribution of matter in large scale structures; 
mass  measurements, whenever feasible, have revealed the presence of 
Dark Matter; galaxies evolution and intercourse with their environment
are responsible for the chemistry of the Universe and ultimately
for life.
X-ray observations  have given us a new key band for understanding 
these building blocks of the Universe, 
with implications ranging from the the study of extreme physical situations, such as 
can be found in the proximity of Black Holes, or near the surface of neutron stars;
to the interaction of galaxies and their environment;
to the measure of parameters of fundamental cosmological importance.

The discrete X-ray source population of galaxies gives us a direct view
of the end-stages of stellar evolution.
Hot gaseous halos are uniquely visible in X-rays. Their discovery in
E and S0 galaxies has
given us a new, potentially very powerful, tool for the measurement
of Dark Matter in galaxies, as well as for local estimates of $\Omega$. 
Galaxy ecology - the study
of the cycling of enriched materials from galaxies into their
environment - is {\em inherently} an X-ray subject. Escape
velocities from galaxies, when thermalized, are kilovolt X-ray
temperatures.
The X-ray band is where we can directly
witness this phenomenon (e.g in M82; NGC~253; see \cite{f89}, \cite{f96}
and refs. therein)(fig~1).

\begin{figure}
  \begin{center}
    \leavevmode
    \includegraphics[height=9.cm,angle=-90.]{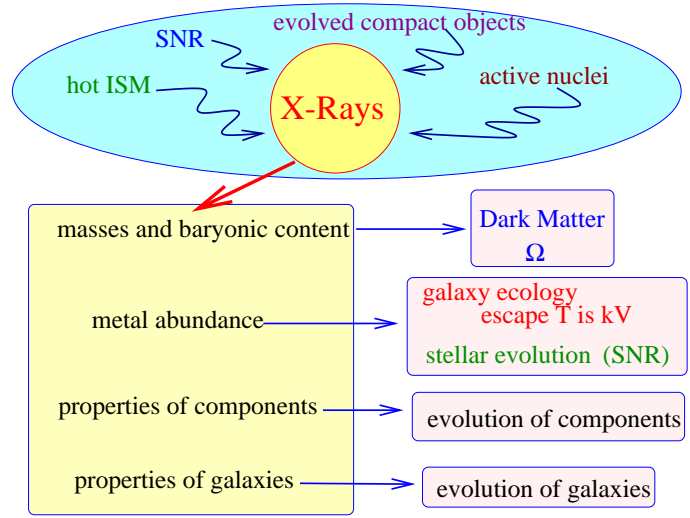}
    \caption{X-ray observations of galaxies and their implications}
  \end{center}
\end{figure}

The study of galaxies and their components in the local universe allows us to establish
the astrophysics of these phenomena. This knowledge can then be used to 
understand the properties of galaxies at the epoch of formation and their
subsequent evolution, both in the field and in clusters.

\section{Requirements for an X-ray Telescope}

Fig.~2 summarizes the requirements for an X-ray telescope that
will significantly advance our knowledge of galaxies. The purpose
of this telescope is two-fold:

\medskip
\noindent
1) {\bf Very detailed
studies} of the X-ray components of nearby galaxies,
to gain the needed deeper astrophysical understanding
of their properties.
Nearby galaxies offer an unique opportunity for studying complete
uniform samples of galaxian X-ray sources (e.g. binaries, SNRs,
black hole candidates),
all at the same distance, and in a variety of environments.
This type of information cannot be obtained for Galactic X-ray
sources, given our position in the Galaxy. These population studies
will be invaluable for constraining X-ray properties and evolution
of different types of sources.

\medskip
\noindent
2) {\bf Study of deep X-ray fields}, where galaxies are
likely to be a very large component of the source population.
Looking back in time, and comparing these results with the detailed
knowledge of the X-ray properties of more nearby objects, we will
be able to study galaxy evolution in the X-ray band. We will be 
able to look at galaxies when substantial outflows were likely
to occur and therefore witness the chemical enrichment of the
Universe at its most critical time.

\begin{figure}
  \begin{center}
    \leavevmode
    \includegraphics[height=9.cm,angle=-90.]{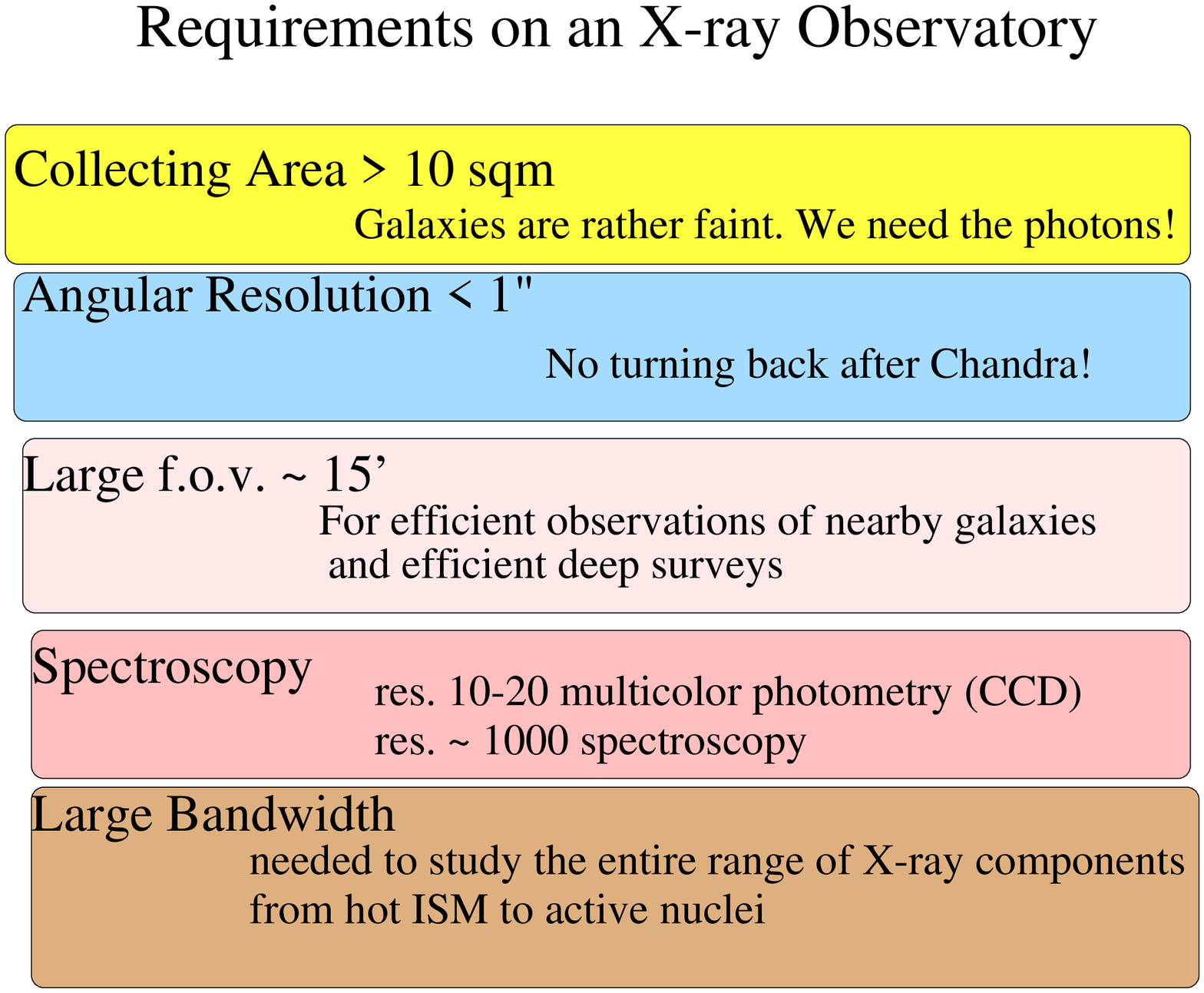}
    \caption{Science driven requirements for a future X-ray telescope}
  \end{center}
\end{figure}

\bigskip
I discuss below in more detail three key elements of fig.~2: collecting area, 
angular resolution, and spectral capabilities.

\medskip
\noindent
{\bf Collecting Area -} A collecting area in the 10-30 sq.meters range is needed for both
in-depth studies of individual galaxies in the nearby Universe
(fig.~3), and for looking back in time (fig.~4).

\begin{figure}
  \begin{center}
    \leavevmode
    \includegraphics[height=9.cm,angle=-90.]{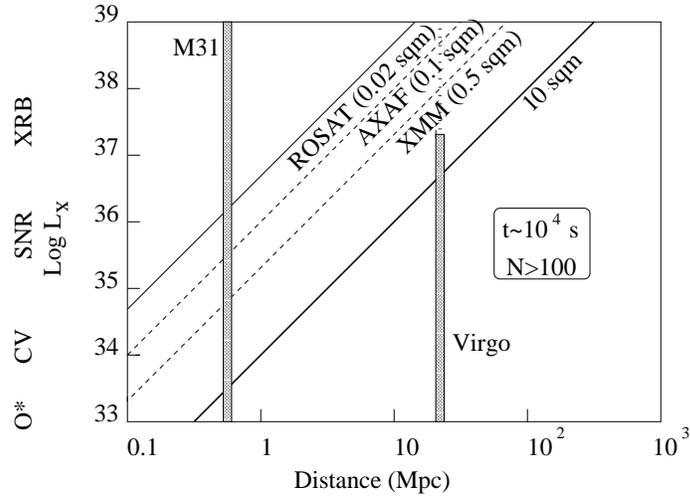}
    \caption{Detection limits of sources in galaxies}
  \end{center}
\end{figure}

\begin{figure}
  \begin{center}
    \leavevmode
    \includegraphics[height=9.cm,angle=-90.]{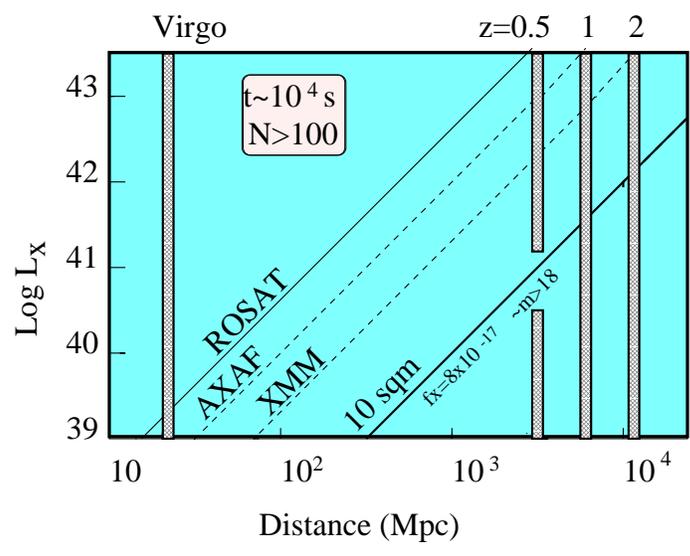}
    \caption{Detection limits of galaxies as a function of distance}
  \end{center}
\end{figure}

\medskip
\noindent
{\bf Angular Resolution -} Arcsecond or better angular resolution is a must, to avoid 
confusion in both the study of nearby galaxies, and in the 
study of deep fields. Chandra images demonstrate the
richness of detail one obtains with subarcsec resolution.
With Chandra-like angular resolution
galaxies can be picked out easily from unresolved stellar-like 
objects in deep exposures. Fig.~5 shows the deep X-ray count that can 
be reached with a 25sqm telescope in
100ks. In the deepest decade galaxies will be a major contributor
and may even dominate the counts, if there is luminosity evolution
in the X-rays, comparable to that observed in the FIR.
Based on the HDF results (\cite{gh}) high z galaxies may be visible.
However, arcsec resolution is needed to avoid confusion at these
faint fluxes (\cite{ef96}; as demonstrated by recent simulations by
G. Hasinger).

Such deep exposures 
would allow the study of the evolution of galaxies in X-rays,
and of the evolution of their stellar binary population as well as of their
hot gaseous component. 
Based on the Madau cosmic SFR, White \& Ghosh (\cite{wg}) show that a comparison of the z-dependence
of the X-ray and optical luminosity functions is related to the evolution
of the X-ray Binary population in galaxies. Moreover,
if hot outflows are prominent at early
epochs we will have a first hand account of the metal
enrichment of the Universe.

\begin{figure}
  \begin{center}
    \leavevmode
    \includegraphics[height=8.5cm,angle=-90.]{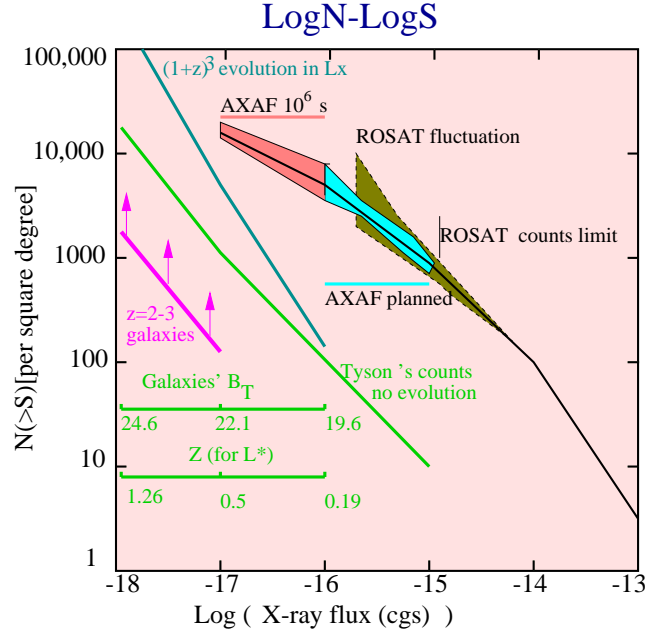}
    \caption{Deep X-ray count that can be reached with a 25sqm telescope in
100ks. }
  \end{center}
\end{figure}

\medskip
\noindent
{\bf Spectral Capabilities -} Fig~6 illustrates the scope of the spectral work one would like
to perform. With X-ray spectroscopy we can determine the
physical status of hot plasmas as well as their chemical composition.
We can also measure cooler ISM by studying the absorption
spectra of background quasars. Spectroscopy goals related to galaxy 
studies are described in figs.~7 and 8.

\begin{figure}
  \begin{center}
    \leavevmode
    \includegraphics[height=9.cm,angle=-90.]{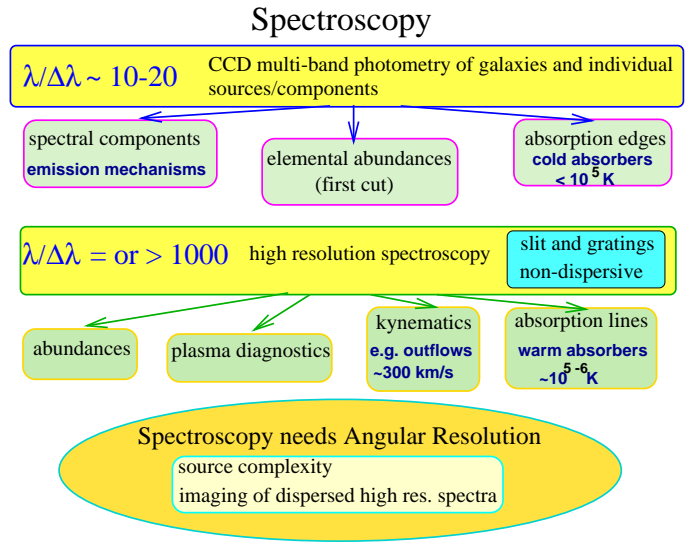}
    \caption{Spectral Capabilities}
  \end{center}
\end{figure}

\begin{figure}
  \begin{center}
    \leavevmode
    \includegraphics[height=9.cm,angle=-90.]{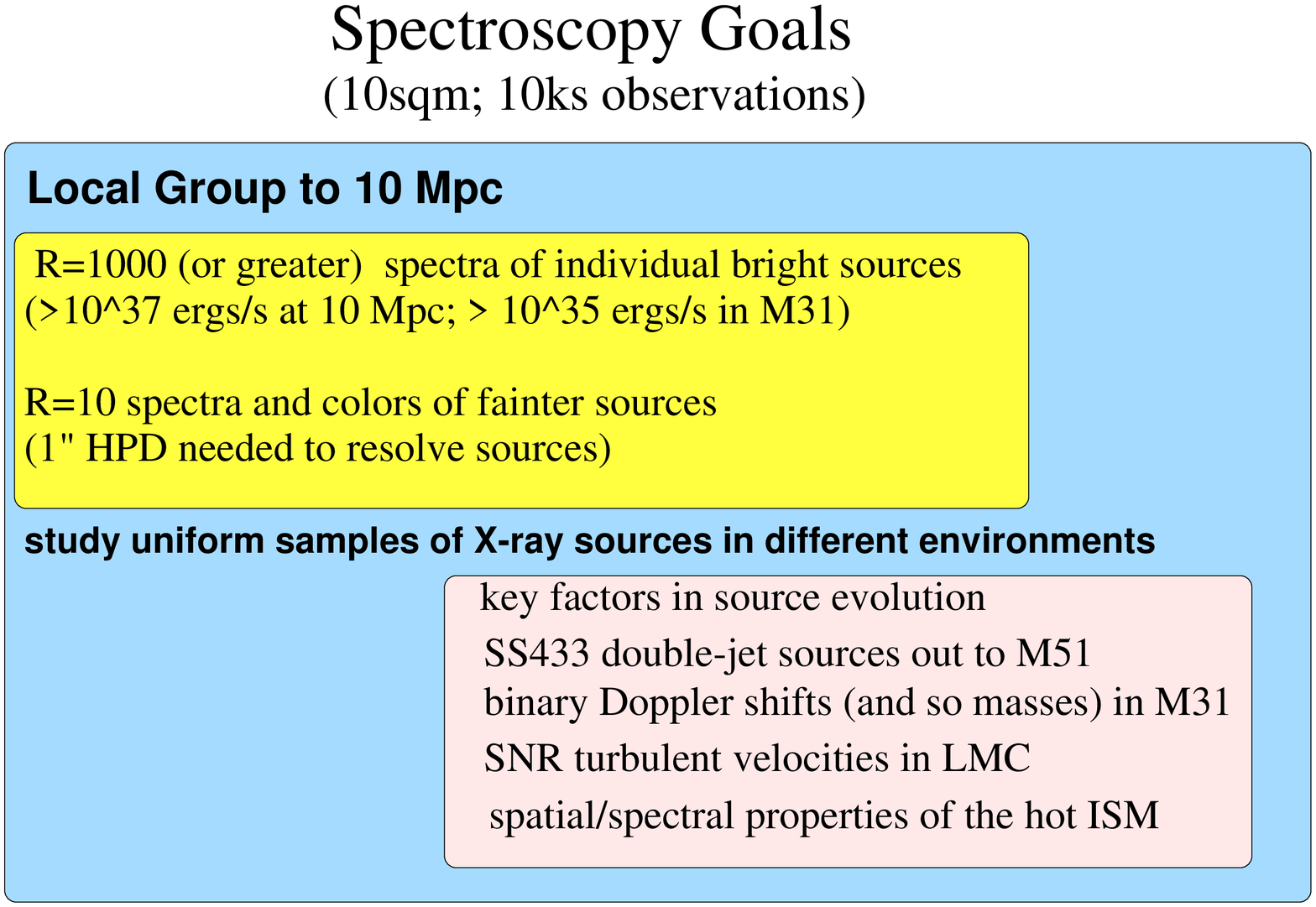}
    \caption{Spectroscopy Goals in the near Universe}
  \end{center}
\end{figure}

\begin{figure}
  \begin{center}
    \leavevmode
    \includegraphics[height=9.cm,angle=-90.]{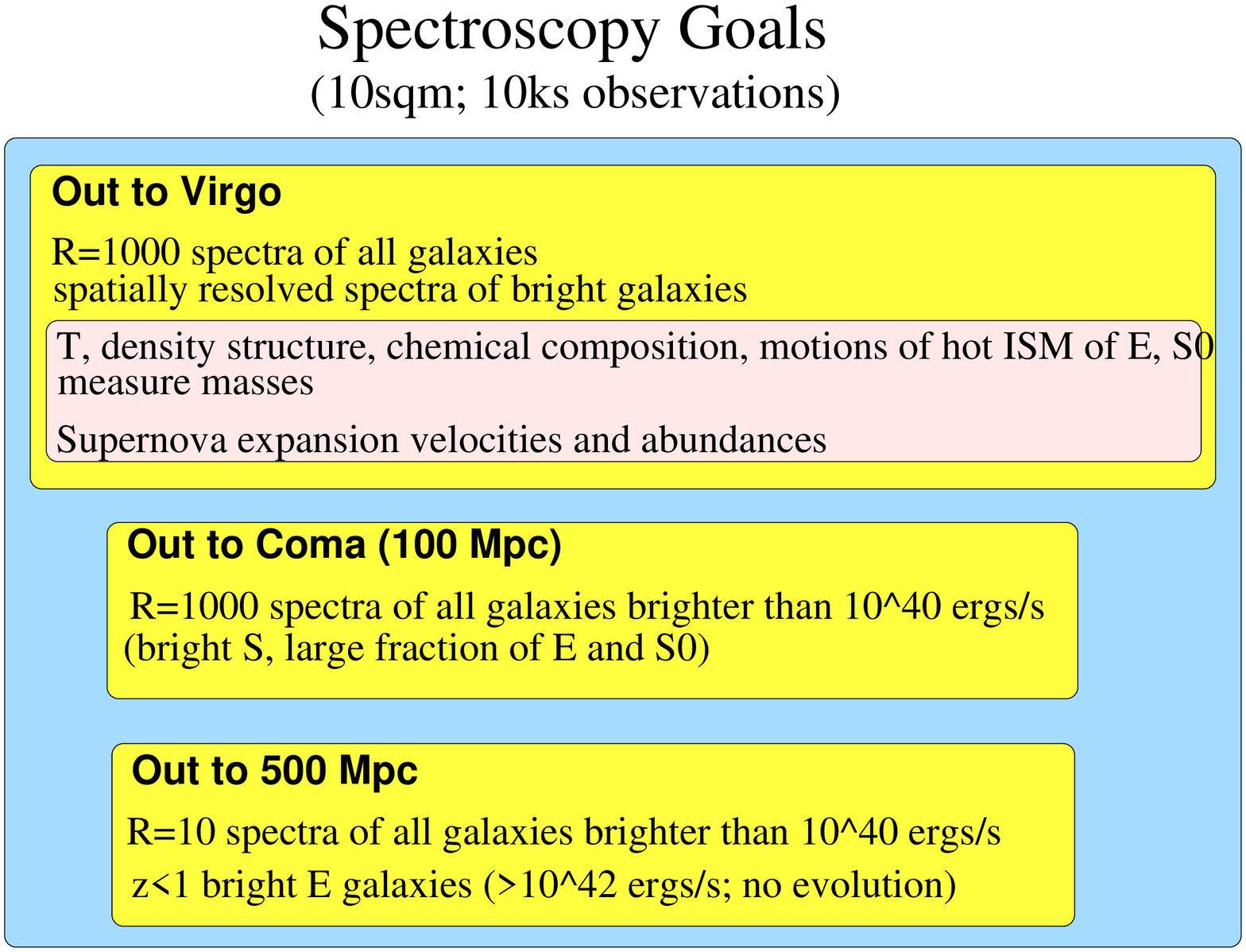}
    \caption{Spectroscopy Goals at larger distances}
  \end{center}
\end{figure}

\bigskip
How do these requirements compare to the characteristics of planned
future X-ray observatories (Constellation X under study by the NASA
community, and XEUS under study in Europe) is shown in Figs~9
and 10. While spectra and bandwidth characteristics of these missions
under study accomplish our goals in both cases, the other requirements
fall below those we need. XEUS has the required large collecting area,
while Con-X area is significantly smaller. In both cases angular 
resolution is significantly sub-Chandra. Based on what Chandra has 
shown and on the characteristics of the objects we want to study
- galaxies are complex objects - a Chandra-like resolution is 
a must. The field of view is also small in both cases, especially
in the case of Con-X.

As we have done in the past \cite{ef96} we advocate that the X-ray 
community consider
a large area, Chandra-like resolution mission to push X-ray astronomy
from an exploratory discipline to a discipline at a par with 
the other wavelength astronomies. Both the scientific potential
of the studies that can be performed with such a telescope,
and more directly the exciting discoveries resulting from
Chandra's high resolution images, support this project.

\begin{figure}
  \begin{center}
    \leavevmode
    \includegraphics[height=10.cm,angle=-90.]{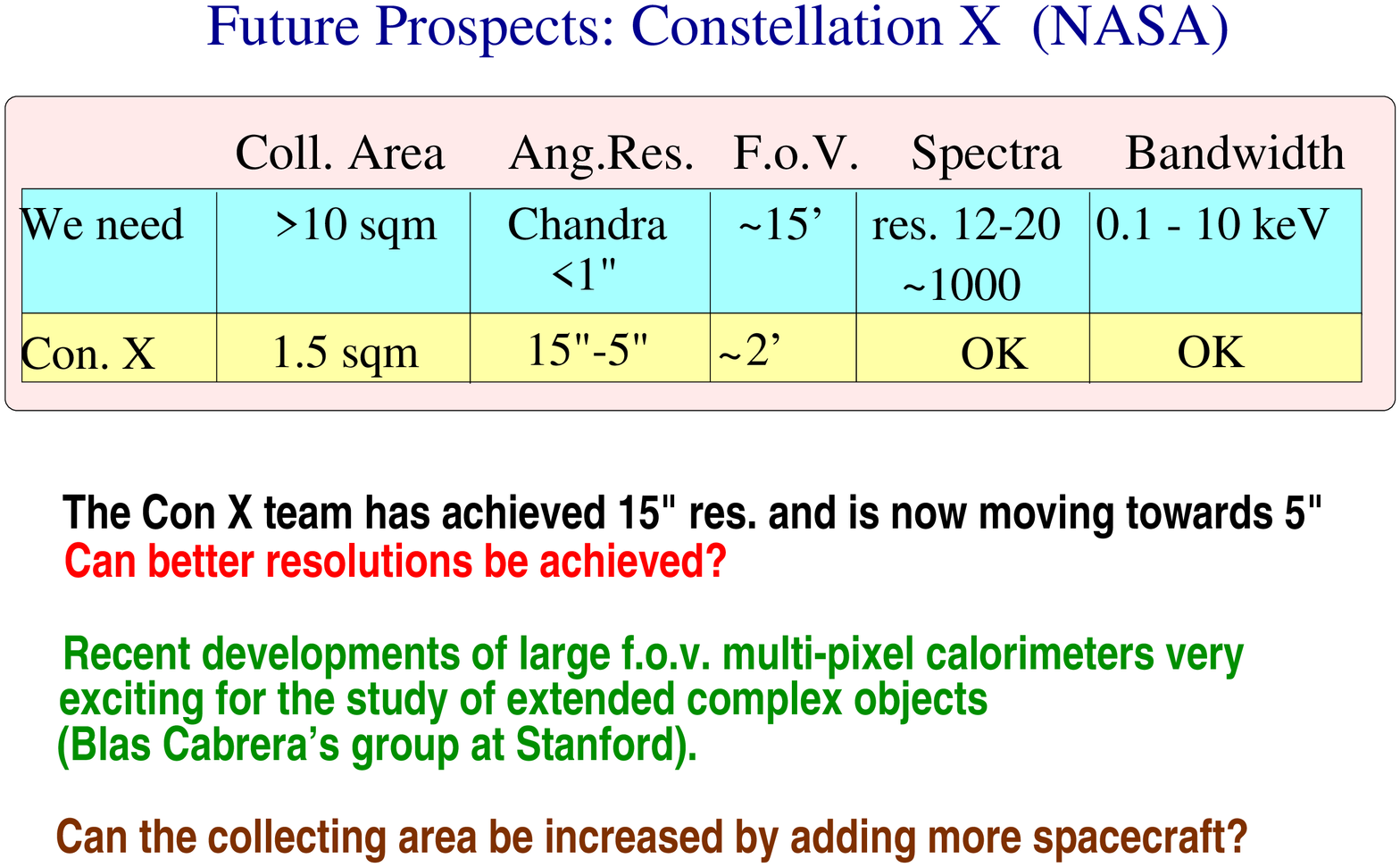}
    \caption{Comparison of our goals with the Constellation X plans}
  \end{center}
\end{figure}

\begin{figure}
  \begin{center}
    \leavevmode
    \includegraphics[height=10.cm,angle=-90.]{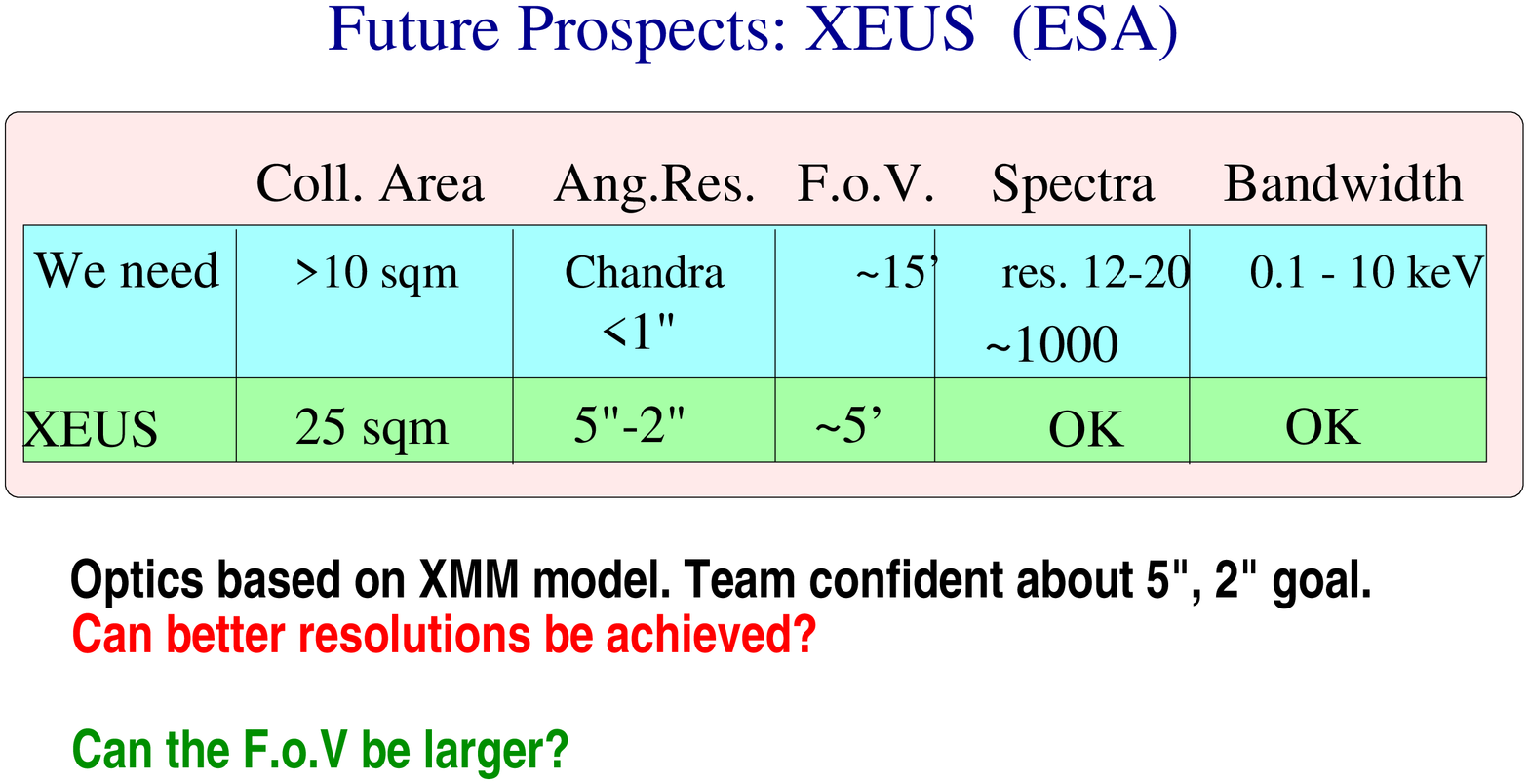}
    \caption{Comparison of our goals with the XEUS plans}
  \end{center}
\end{figure}

\vfill

\acknowledgements{I acknowledge partial travel support from the Chandra Science Center.
Parts of this talk were also given at the Cosmic Genesis Workshop held at Sonoma
State University, Oct 27-30, 1999, and at the meeting Astrophysical Plasmas: 
Codes, Models \& Observations, Oct 25-29, Mexico City, and will be included
in the proceedings of these meetings.
Part of the material presented here has also been presented at the XEUS
Symposium 1999. I thank my colleagues
who have contributed thoughts and material included in this paper. In 
particular, I acknowledge fruitful discussions with Martin Elvis,
Pat Slane, Josh Grindlay and Paul Gorenstein.}
\vfill
\end{document}